# Robust full-pose-parameter estimation for the LED array in Fourier ptychographic microscopy

CHUANJIAN ZHENG,[1] SHAOHUI ZHANG,[1, *] DELONG YANG,[1] GUOCHENG ZHOU,[1] YAO HU,[1] AND QUN HAO[1]

[1]*School of Optics and Photonics, Beijing Institute of Technology, Beijing 100081, China*
*zhangshaohui@bit.edu.cn*

**Abstract:** Fourier ptychographic microscopy (FPM) can achieve quantitative phase imaging with a large space-bandwidth product by synthesizing a set of low-resolution intensity images captured under angularly varying illuminations. Determining accurate illumination angles is critical because the consistency between actual systematic parameters and those used in the recovery algorithm is essential for high-quality imaging. This paper presents a full-pose-parameter and physics-based method for calibrating illumination angles. Using a physics-based model constructed with general knowledge of the employed microscope and the brightfield-to-darkfield boundaries inside captured images, we can solve for the full-pose parameters of misplaced LED array, which consist of the distance between the sample and the LED array, two orthogonal lateral shifts, one in-plane rotation angle, and two tilt angles, to correct illumination angles precisely. The feasibility and effectiveness of the proposed method for recovering random or remarkable pose parameters have been demonstrated by both qualitative and quantitative experiments. Due to the completeness of the pose parameters, the clarity of the physical model, and the high robustness for arbitrary misalignments, our method can significantly facilitate the design, implementation, and application of concise and robust FPM platforms.

## 1. Introduction

Fourier ptychographic microscopy (FPM) [1-3] is a recently developed computational imaging technique. It uses angle-varied illumination to surpass the resolution limit in optical imaging systems. By integrating synthetic aperture [4] and phase retrieval [5] concepts, the space-bandwidth product (SBP) of imaging the complex transmittance function characterizing the absorption and phase modulation properties of samples can be significantly improved. Because of its superior performance, FPM has been a powerful technique in many fields, such as digital pathology [6], aberration metrology [7], high-throughput cytometry [8], and three-dimensional (3D) imaging [9-11], among the others.

In contrary to real-space ptychography [12, 13], FPM avoids the use of mechanical scanning devices and instead utilizes a programmable LED array to illuminate a sample of interest from multiple angles, thereby enabling aperture scanning in the Fourier domain. At each illumination angle, FPM captures a low-resolution (LR) intensity image, corresponding to a circular sub-spectrum whose position and radius are determined by the illumination wavelength, illumination angle, and the numerical aperture (NA) of the objective. Then, the FPM algorithm can recover high-resolution (HR) complex sample images by sequentially imposing amplitude constraint with each captured image in the spatial domain, and support region constraint with associate circular sub-spectrum in the Fourier domain.

As a typical computational imaging method, FPM has high requirements for the consistency between the illumination parameters of the experimental LED array and those used in the recovery algorithm. Nevertheless, when one builds or modifies an FPM platform, the LED array is unavoidable to be misaligned, resulting in inaccurate stitching of the LR images' spectra in the Fourier domain, in turn, degrading the quality of the recovered images. Two types of methods are proposed to overcome the problem, which are pre-calibration and post-calibration, respectively. For the pre-calibrated scheme, one can use precision mechanical devices to adjust the LED array to an ideal position before capturing the raw data set based on the symmetrical feature of optical systems [14, 15], or utilize the shift of captured out-of-focus image to seek the correct illumination angles [16]. One can also obtain the location and orientation of the LED array by exploiting the features of brightfield-to-darkfield (B-D) boundaries [17]. However, these methods not only need significant user expertise or specific devices, but also are not robust to changes in the system (e.g., bumping the setup). To alleviate the cost- and labor-intensive task, many post-calibrated methods are developed. For example, pcFPM [18] and SC-FPM [19] can obtain four non-titled pose parameters of the LED array by using the intensity distribution of captured images, and another two-part calibration [20] can also achieve the purpose using the spectra of images. Unfortunately, SC-FPM and pcFPM are based on simulated anneal (SA) algorithm, the performance of which highly relies on suitable initialization of step size and search direction. When the LED is remarkably misplaced, one has to manually refine the initial parameters to avoid the optimization algorithm falling into a local trap or failing to converge. Furthermore, all post-calibrated methods have two limitations. Firstly, they are not robust because the captured data is coupled with not only the pose misalignment, but also the aberration of objective, intensity fluctuation, and system noise. Secondly, they are not efficient because a number of iterative times are needed for the convergence of estimated pose parameters.

To combine the advantages of pre-calibrated and post-calibrated approaches, and obtain the pose parameters of all 6 degrees of freedom of the LED array, we report a physics-based full-pose-parameter estimation method for FPM platforms. With the general knowledge of the employed microscope and captured LR images, our method can solve for the six pose parameters to model the LED array precisely for automatically calibrating illumination angles. We show the principle of forming an arc-shaped B-D boundary inside the captured image, and construct an explicit physics model to connect the position of each LED element with the circle center and radius of the B-D boundary. In addition, to extract the B-D boundary from a captured image, we design an algorithm to remove noise and accurately calculate the circle center and radius of the B-D boundary. The feasibility and effectiveness of calibrating pose misalignment have been demonstrated by experiments. Such a physics-based method is capable of obtaining the full-pose parameters of LED array placed randomly, significantly prompting the development of adjust-free FPM platforms.

The remainder of this paper is organized as follows: we review the theory of FPM in Section 2.1. Then we introduce a precise LED array model and analyze the misalignment-induced artifacts in Section 2.2. In Section 2.3, we construct a physics model to connect the position of LEDs with captured B-D boundary features. In Section 2.4, we proposed a full-pose-parameter estimation method to correct pose misalignment based on the constructed model. Experimental results with resolution target and biological sample are presented in Section3. Finally, we summarize the conclusions in Section 4.

## 2. Principle

### 2.1 FPM theory

A typical FPM system is composed of an LED array, a low-NA objective, a tube lens, and an image sensor (COMS or CCD camera). Experimentally, M × N LEDs are turned on sequentially to illuminate a thin sample that is placed far away from the LED array, and the image sensor is used to record the corresponding LR images. For a small segment of the sample, whose size is sufficient small compared with the distance between the LED array and the sample, the light wave from $LED_{m,n}$ (row $m$, column $n$) can be approximately treated as a plane wave with wave vectors of $(u_{m,n}, v_{m,n})$. The LR intensity image can be described as

$$I_{m,n}^c(x, y) = |[o(x,y)e^{(jxu_{m,n}, jyv_{m,n})}] * h(x,y)|^2 = |\mathcal{F}^{-1}\{O(u-u_{m,n}, v-v_{m,n}) \cdot P(u,v)\}|^2, \quad (1)$$

where $o(x, y)$ is the complex transmission function of the thin sample, $(x, y)$ denotes the two-dimensional (2D) coordinates in the sample plane, $j$ is the imaginary unit, $h(x, y)$ stands for the point spread function, $*$ is the convolution operator, and $\mathcal{F}^{-1}$ represents the inverse Fourier transform operator. $O(u,v) = \mathcal{F}\{o(x, y)\}$ and $P(u,v) = \mathcal{F}\{h(x, y)\}$ are the Fourier transform of $o(x, y)$ and $h(x, y)$, respectively. For a circularly-symmetrical diffraction-limited coherent imaging system the transfer function, $P$, is a circular pupil determining the highest spatial frequency transmitted by the objective as NA/λ, where λ is the central illumination wavelength.

FPM computationally generates HR complex images from the captured M × N LR images with a recovery procedure composed of five steps. The first step is to initialize the guesses of the HR sample spectrum $O_0(u,v)$ and pupil function $P_0(u,v)$, where the subscript '0' denotes the index of initial iteration. In general, the initial spectrum is set as the Fourier transform of any up-sampled LR image. The initial pupil function is given as a circular low-pass filter with one and zero amplitudes within and outside the passband, and zero phase everywhere. Secondly, use the pupil function and spectrum to generate a LR target image under the illumination of LED$_{m,n}$ as

$$o^e_{0,m,n}(x, y) = \mathcal{F}^{-1}\{O_0(u - u_{m,n}, v - v_{m,n})P_0(u,v)\}. \tag{2}$$

Thirdly, replace the target image's amplitude components with the square root of the intensity image obtained under the corresponding illumination angle, and keep the phase unchanged to form an updated LR target image as

$$o^u_{0,m,n}(x, y) = \sqrt{I^c_{m,n}(x, y)} \frac{o^e_{0,m,n}(x, y)}{|o^e_{0,m,n}(x, y)|}. \tag{3}$$

Subsequently, the updated target image is utilized to update the corresponding sub-spectrum of the HR sample spectrum, which is given by:

$$\begin{aligned} O_{i+1}(u - u_{m,n}, v - v_{m,n}) &= O_i(u - u_{m,n}, v - v_{m,n}) + \alpha \frac{P_i^*(u,v)}{|P_i(u,v)|^2_{\max}} \Delta O_{i,m,n}, \\ P_{i+1}(u,v) &= P_i(u,v) + \beta \frac{O_i^*(u - u_{m,n}, v - v_{m,n})}{|O_i(u - u_{m,n}, v - v_{m,n})|^2_{\max}} \Delta O_{i,m,n}, \end{aligned} \tag{4}$$

where $\alpha$ and $\beta$ are the iterative step size, which can be set as one or adaptively updated to decrease the noise in recovered image [21]. The subscript $i$ denotes the index of iteration, and $\Delta O_{i,m,n} = \mathcal{F}\{o^u_{i,m,n}(x, y)\} - \mathcal{F}\{o^e_{i,m,n}(x, y)\}$ is an auxiliary function.

In the fourth step, steps 2-3 are repeated for other captured LR images. Finally, steps 2-4 are repeated until the solution converges. At the end of the iterative recovery process, the converged solution in the Fourier domain will cover a significantly extended spectral support, and it will be transformed to the spatial domain to recover HR intensity and phase images.

### 2.2 LED array model and misalignment-induced artifacts

It is worth noting that the image recovery in FPM is based on the accurate stitching of each sub-spectrum in the Fourier domain, as detailed in the fourth step of the recovery procedure. Indeed, the position of each LED element with respect to the sample is critical for a successful recovery, as it directly determines the center coordinate of the stitched sub-spectrum. In practice, LEDs are arranged on a plane board in a matrix way. Both the plane and matrix distribution are key prior constraints to determine the full-pose parameters of the LED array in our method.

A global LED array model [19] has been proposed to constrain the position of LEDs to ensure the convergence of SA algorithm. It uses four parameters ($\Delta x$, $\Delta y$, $\theta_z$, $h$) to model an LED array, where $\Delta x$ and $\Delta y$ are the lateral shifts, $\theta_z$ is the in-plane rotation angle, and $h$ is the distance between the sample and the LED array. However, the pose of a rigid LED board in the 3D space should be described with the pose parameters of 6 degrees of freedom. Therefore, we improve the model with more complete pose parameters ($\Delta x$, $\Delta y$, $\theta_x$, $\theta_y$, $\theta_z$, $h$), where $\theta_x$ and $\theta_y$ are the tilt angles along the $x$- and $y$-axes, respectively. Fig. 1 illustrates the LED array modeled with the six parameters. A programmable LED array (5×5 LEDs are sketched for the sake of clarity) is placed under the sample. The coordinate system is built as follows: the $z$-axis is the optical axis of the objective with a direction from the LED array to the sample, and it interacts with the LED array at the point $O$. The $y$-axis passing through the point $O$ has the same direction as the line that goes through the centers of two adjacent pixels on the image sensor. The $x$-axis also passes through the point $O$ and is perpendicular to the y-axis.

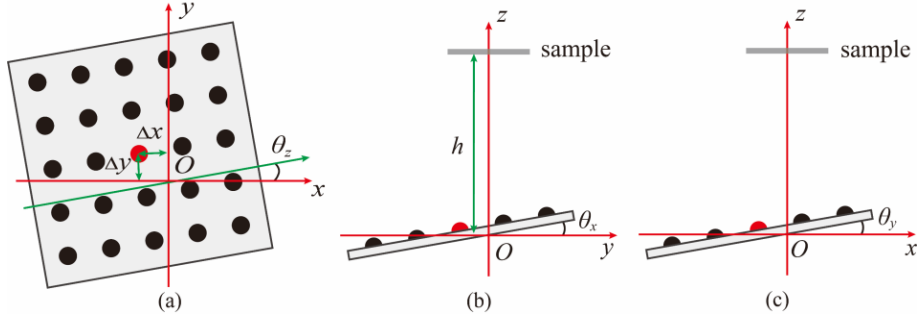

Fig.1. Schematic of a misaligned LED array model. (a) shows the misaligned LED array with lateral shifts of ($\Delta x$, $\Delta y$) and the in-plane rotation angle of $\theta_z$. (b)-(c) are the side views of the misaligned LED array, where $\theta_x$ and $\theta_y$ are the tilts angles along $x$- and $y$-axes, respectively.

We assume the distance between adjacent LEDs is the same, then the position of each individual LED element can be expressed as

$$\begin{bmatrix} x_{m,n} \\ y_{m,n} \\ z_{m,n} \end{bmatrix} = \mathbf{R} \begin{bmatrix} md_{LED} \\ nd_{LED} \\ 0 \end{bmatrix} + \begin{bmatrix} \Delta x \\ \Delta y \\ 0 \end{bmatrix}, \tag{5}$$

where ($x_{m,n}$, $y_{m,n}$, $z_{m,n}$) represents the coordinate of $LED_{m,n}$ in the 3D space, $d_{LED}$ denotes the distance between adjacent LEDs, and $\mathbf{R}$ is the rotation matrix that can be written as

$$\mathbf{R} = \begin{bmatrix} \cos(\theta_y)\cos(\theta_z) & \cos(\theta_y)\sin(\theta_z) & -\sin(\theta_y) \\ -\cos(\theta_x)\cos(\theta_z) + \sin(\theta_x)\sin(\theta_y)\cos(\theta_z) & \cos(\theta_x)\cos(\theta_z) + \sin(\theta_x)\sin(\theta_y)\sin(\theta_z) & \sin(\theta_x)\cos(\theta_y) \\ \sin(\theta_x)\sin(\theta_z) + \cos(\theta_x)\sin(\theta_y)\cos(\theta_z) & -\sin(\theta_x)\sin(\theta_z) + \cos(\theta_x)\sin(\theta_y)\sin(\theta_z) & \cos(\theta_x)\cos(\theta_y) \end{bmatrix}. \tag{6}$$

Once we obtain the coordinate of each LED, we can calculate the illumination wave vectors as

$$u_{m,n} = \frac{2\pi}{\lambda} \cdot \frac{x_c - x_{m,n}}{\sqrt{(x - x_{m,n})^2 + (y - y_{m,n})^2 + (h - z_{m,n})^2}},$$

$$v_{m,n} = \frac{2\pi}{\lambda} \cdot \frac{y_c - y_{m,n}}{\sqrt{(x - x_{m,n})^2 + (y - y_{m,n})^2 + (h - z_{m,n})^2}},$$

(7)

where $(x_c, y_c)$ is the central coordinate of one small segment of the sample.

To illustrate the misalignment-induced artifacts and demonstrate the necessity of pose calibration, we perform simulations and compare the effects of different pose parameters quantitatively. The simulation parameters are chosen based on an experimental system. A 21 × 21 LED array (2.5 mm LED spacing and 470 nm wavelength) is placed 92 mm beneath the sample. The NA and magnification of the objective are 0.1 and 4×, respectively. Figs. 2(a1) and 2(a2) are the ground truth of the sample. First, we study the effect of lateral-shift-induced artifacts by increasing the value of $\Delta x$ from 0 μm to 2000 μm with a step size of 100 μm. The curves of root-mean-square error (RMSE) between the recovered images and the ground truth are shown in Figs. 2(b3) and 2(b4), where $I$ and $\Phi$ represent the intensity and phase, respectively. When $\Delta x$ varies from 600 μm to 700 μm, the RMSE of recovered intensity and phase images increases drastically, and the artifacts arising in the images degrade the imaging quality severely, as shown in Figs. 2(b1) and 2(b2). Therefore, calibrating the lateral shift is necessary for high-quality imaging.

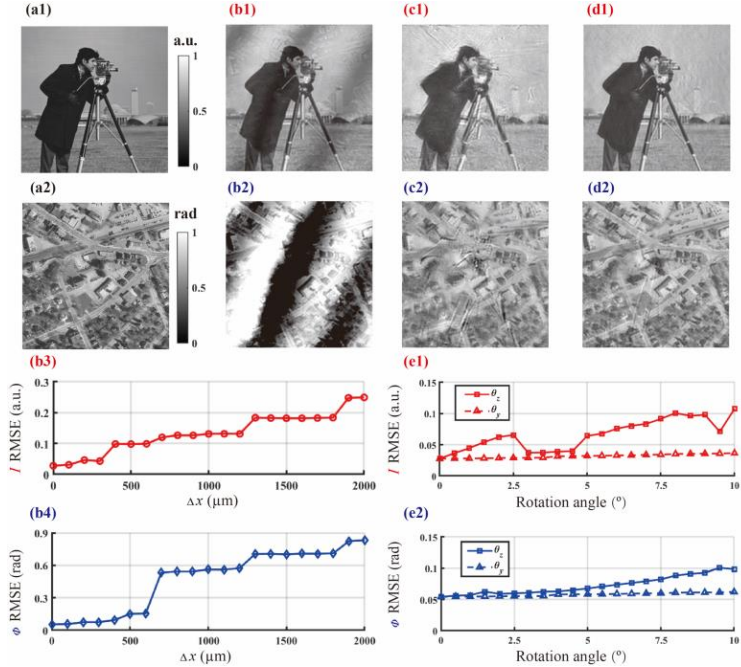

Fig.2. Effects of different pose misalignments. (a1)-(a2) are the ground truth of the sample; (b1)-(b2) are the recovered intensity and phase images, respectively, with a lateral shift of 700 μm along the $x$ axis ($\Delta x$=700 μm). (c1)-(c2) are the recovered intensity and phase images, respectively, with an in-plane rotation angle of 10° along the $z$ axis ($\theta_z$=10°). (d1)-(d2) are the recovered intensity and phase images, respectively, with a tilt angle of 10° along the $y$ axis ($\theta_y$=10°); (b3)-(b4) show the RMSE between the recovered images ($I$ is intensity, and $\Phi$ is phase) and the ground truth with increased lateral shift. (e1)-(e2) show the RMSE between the recovered images ($I$ and $\Phi$) and the ground truth with increased in-plane rotation ($\theta_z$) and tilt angle ($\theta_y$), respectively.

Next, we analyze the effect of in-plane rotation and tilt with similar simulations. The curves of RMSE with increased rotation angle are shown in Figs. 2(e1) and 2(e2), respectively. We find that the RMSE induced by either in-plane rotation or tilt is smaller than that induced by lateral shift. Although the rotation angle increases to 10°, which is easy to be observed only with the human eye, the distortion in recovered images is still slight, as shown in Figs. 2(c) and 2(d). The different effects can be interpreted with two simple equations. For lateral shift, all LEDs are shifted by $\Delta x$, resulting in the overall shift of spectra, whatever the low or high frequencies. However, the shifts of the position of LED$_{m,n}$ caused by in-plane rotation are $x_{m,n}[1-\cos(\theta_z)]$ and $x_{m,n}\sin(\theta_z)$ along the *x*- and *y*- axes, respectively, and the tilt-induced shift is also $x_{m,n}[1-\cos(\theta_y)]$ along the *x*-axis. Therefore, the shifts caused by rotation or tilt are linear to the coordinate of the LED element; that is, the shift of the spectrum with low-frequency information is smaller than that with high-frequency information. Meanwhile, the low-frequency information of an image is usually more important and easier to be observed. That is why the degradations of image quality of Figs. 2(c) and 2(d) are much smaller than that of Figs. 2(b1) and 2(b2). In any case, both in-plane rotation and tilt have degraded the image quality, and thus they are needed to be calibrated for accurate quantitative analysis in FPM.

*2.3 Physics-based model for pose calibration*

In computational imaging approaches, an important feature is to seek the joint optimization of imaging systems and algorithms. One of the core ways to achieve that is to mine and utilize more effective and explicit physical models. To calibrate the pose misalignment, we construct a physics-based model to connect the B-D boundary inside captured image with the full-pose parameters.

As shown in Fig. 3(a1), when the central LED element is used to provide illumination, the diameter of the light is greater than the FOV of the CMOS camera, so the captured intensity image is a pure brightfield (BF) image. Fig. 3(b1) shows such a BF image, where the circle drawn by the green dashed line is the projection of the aperture stop, point $C_{0,0}$ is the circle center, and the red rectangle is the FOV of the CMOS camera that serves as the field stop of the system. In contrast, the intensity image with the illumination of the LED element whose illumination NA is close to the NA is a B-D image, as sketched in Fig. 3(b2). Typically, the B-D boundary is arc-shaped because it is formed by the superposition of the field stop and the projection of the circular aperture stop in the image plane, as shown in Fig. 3(a2). In order to analyze conveniently, we transform the field stop from the image space into the object space and get the entrance window (EW), and transform the aperture stop and get the entrance pupil (EP). The schematic diagram of our physics-based model is presented in Fig.3 (c).

To model mathematically, we first calculate the distance between the EP and EW, denoted by $h_1$. It is noted that $h_1$ is a constant characteristic quantity of an FPM system because the position of EP and EW are fixed. As shown in Fig. 3(c), *GH* is the EP's radius calculated as $h \cdot NA$. $CC_{0,0}$ is the radius of the projection of the EP on the EW plane. We note that the projection is conjugated to the B-D boundary according to the above transformation, so $CC_{0,0}$ can be calculated as $R_{BD}$/Mag, where $R_{BD}$ is the radius of the B-D boundary, and Mag is the magnification of the system. With a simple geometrical deduction, we can get

$$\frac{CC_{0,0}}{GH} = \frac{h}{h-h_1}. \tag{8}$$

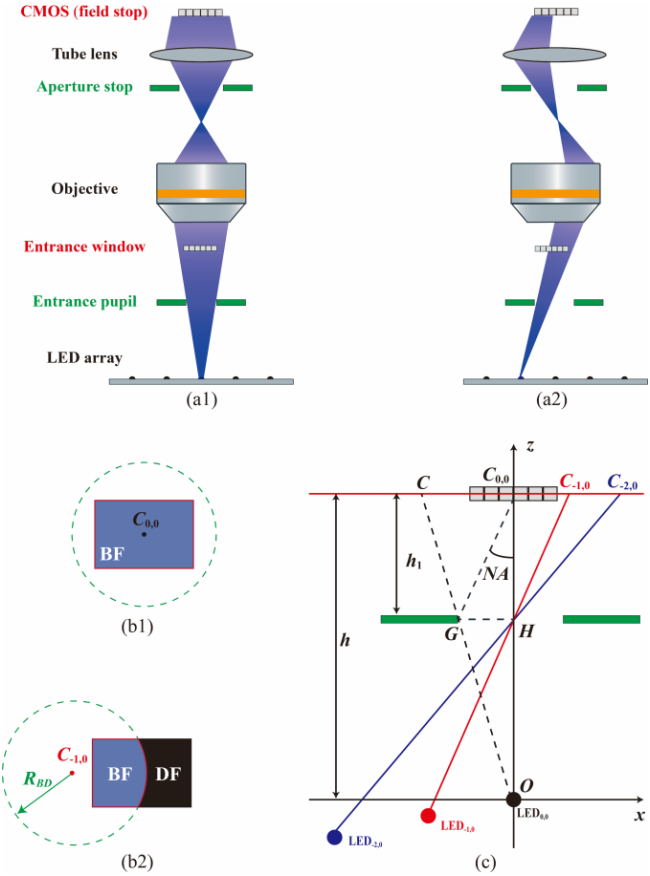

Fig.3. Physics-based model used for pose calibration in FPM system. (a1) shows the principle of forming a BF image, and the corresponding image is shown in (b1). (a2) shows the principle of a B-D image, and the corresponding image is shown in (b2). (c) is the physics-based model used to connect the features of B-D image with the position of each LED element.

Substituting the formulas of $CC_{0,0}$ and $GH$ into Eq. (8), we can get

$$h_1 = \frac{hR_{BD}}{\mathrm{NA}\cdot h + R_{BD}}. \tag{9}$$

Now, we can calculate the circle center of the B-D boundary as

$$\begin{aligned}x_{BD,m,n} &= \frac{h_1 \cdot x_{m,n}}{h - h_1 - z_{m,n}},\\ y_{BD,m,n} &= \frac{h_1 \cdot y_{m,n}}{h - h_1 - z_{m,n}}.\end{aligned} \tag{10}$$

Eqs. (9) and (10) are the forward model of our method, which suggest that we can calculate the center circle and radius of a B-D boundary with the knowledge of the position of corresponding LED element and the NA of the objective. In turn, we can calculate the position of each LED with the features of B-D boundary to further solve for the full-pose parameters. More details are shown in the following Section.

### 2.4 Physics-based and self-calibrated strategy

Unlike existing pre- or post-calibrated methods, we present a physics-based full-pose-parameters calibration method. It takes full advantage of general knowledge of the employed microscope, and can achieve robust and precise pose calibration with the features of B-D boundaries inside captured images. The flow chart of our method is shown in Fig. 4.

First, capture LR intensity images as the raw data set. This step is multiplexed with the raw data acquisition of FPM, because the proposed method also belongs to a post-calibrated method. The difference is that the LED array used in our method can be placed randomly.

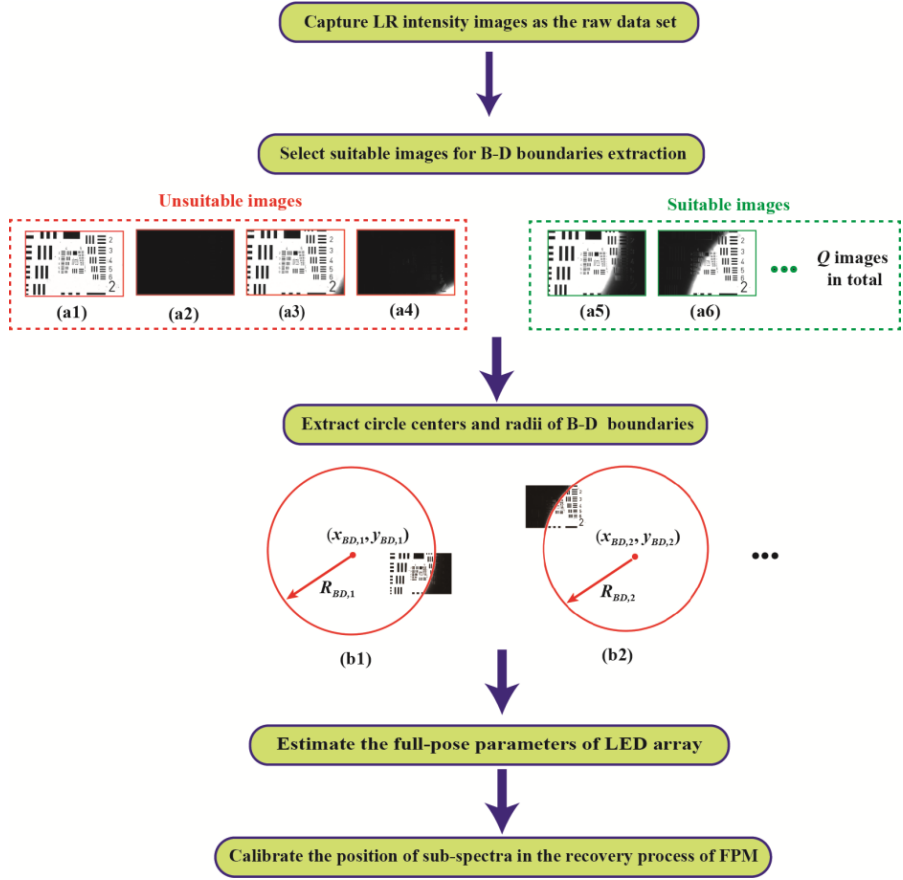

Fig.4. Flow chart of our method. (a1)-(a4) are four unsuitable intensity images for extracting the B-D boundary due to pure BF imaging, pure DF imaging, high-proportion BF imaging, and low-proportion BF imaging. (a5)-(a6) are two suitable intensity images, and the extracted circle centers and radii are depicted in (b1) and (b2).

Second, select suitable images for B-D boundaries extraction. As mentioned in Section 2.3, the position of each LED element can be inferred from the B-D boundary. However, not all captured images can carry suitable B-D information. If an LED element is close to the optical axis, the captured image will be a pure BF image, as shown in Fig. 4(a1). If the illumination NA is greater than the objective's NA, the full FOV of the captured image is DF imaging, as shown in Fig. 4(a2). Extracting B-D information from such images is useless. Thus, we define the extracting range of intensity images as $S=\{(m,n)\,|\,m=-m_{BD},...,m_{BD}, n=-n_{BD},...,n_{BD}\}$, where $m_{BD}$ is the same as $n_{BD}$ for uniformity. In actual experiments, $m_{BD}$ can be determined according to the index of the captured image where the full FOV starts to become DF imaging. Alternatively, it can also be mathematically calculated as

$$m_{BD} = \left\lceil \frac{h \cdot \tan(\arcsin(NA)) + \frac{0.5L_x}{Mag}}{d_{LED}} \right\rceil, \tag{11}$$

where $\lceil \ \rceil$ rounds numbers to upper integers of their argument, and $L_x$ is the length of the COMS camera. In addition, the accuracy of extraction is influenced by the proportion of the BF region to the whole FOV. For example, the intensity images shown in Figs. 4(a3) and 4(a4) are not suitable for following extraction because the proportion of BF region is too high or too low. Therefore, we further define a proportion threshold $\eta_{th} = 0.85$ to judge whether the intensity images within the extracting range are suitable. For an intensity image $I_{m,n}(x,y)$, the proportion of BF region can be approximately calculated as

$$\eta_{m,n} = \frac{\sum_{x,y}^{X,Y} B_{m,n}(x,y)}{XY}, (m,n) \in S, \tag{12}$$

where $B_{m,n}(x,y)$ is the binarized image of $I_{m,n}(x,y)$, $X$ and $Y$ are the number of pixels along the $x$- and $y$-axes, respectively. If $\eta_{m,n} \in [1-\eta_{th}, \eta_{th}]$, we think that $I_{m,n}(x,y)$ is suitable for the following steps. At the end of this step, we define the pre-selected image set as $I_q(x,y)$, $q = (1,...,Q)$, where the subscript $q$ is the index corresponding to $(m, n)$, and $Q$ is the total number of the selected images.

Third, extract circle centers and radii of B-D boundaries. This step is an image processing problem. But some existing methods, e.g., Hough transform that relies on accurate edge detector, are not appropriate for our purpose, because the captured intensity images coupled with sample information will make edge detection problematic. Therefore, we propose an algorithm to solve the problem. The pipeline of algorithm is shown in Tab. 1. For the $q$th image, we first sequentially perform binarizing, removing sample information, and edge detecting. However, one simple image preprocessing scheme cannot satisfy the need to remove the information of various samples. The left sample information will serve as noise, which will degrade the accuracy of extracting B-D boundaries. To maintain the generalizability of the algorithm, we first perform denoise using the coarse estimates of the B-D boundary with random sample consensus (RANSAC) algorithm and then gain refined estimates with least squares (LS) method. In RANSAC algorithm, the total iteration times are noted as $K$. For the $k$th iteration, random 3 edge points are chosen to calculate a circle with center and radius denoted by $(\hat{x}_{BD,q,k}, \hat{y}_{BD,q,k})$ and $\hat{R}_{BD,q,k}$, respectively. A metric function $F_k = \sum_{x_e, y_e} p(x_e, y_e)$, is also calculated to represent the fitness of the current circle, where $(x_e, y_e)$ donates the coordinate of edge points, $p(x_e, y_e)$ is a function used for judging whether an edge point locates on the circle:

$$p(x_e, y_e) = \begin{cases} 1, & if \ |\sqrt{(x_e - \hat{x}_{BD,q,k})^2 + (y_e - \hat{y}_{BD,q,k})^2} - \hat{R}_{BD,q,k}| < th_1 \\ 0, & otherwise \end{cases} \tag{13}$$

where $th_1$ is a distance threshold. After $K$ iterations, we locate the index of the largest $F_k$ and pick up the corresponding estimates $(\hat{x}_{BD,q}, \hat{y}_{BD,q})$ and $\hat{R}_{BD,q}$ as the output. To denoise, we define a new distance threshold, $th_2$. If the difference between the $\hat{R}_{BD,q}$ and the distance from an edge point to the center, $(\hat{x}_{BD,q}, \hat{y}_{BD,q})$, is greater than $th_2$, we consider the edge point as noise and remove it from the existing point set. After removing all noise points, we use LS method to obtain the refined estimates of circle center and radius, denoted by $(x_{BD,q}, y_{BD,q})$ and $R_{BD,q}$, respectively. Two fitted circles with our algorithm are shown in Figs. 4(b1) and 4(b2).

**Table 1: Pipeline of extracting circle centers and radii of B-D boundaries.**

**Input:** $Q$ preselected intensity images $I_q(x,y), q = 1,...,Q$
**Output:** The circle centers $(x_{BD,q}, y_{BD,q})$ and radii $R_{BD,q}$ of the B-D boundaries

for $q = 1$ to $Q$ do
    **Step 1:** Image preprocessing
1:   $B_{q,B-D,sam}(x,y) \leftarrow$ binarize ($I_q(x,y)$, 0.8)    ▶ binarize $I_q(x,y)$ with a threshold of 0.8
2:   $B_{q,B-D}(x,y) \leftarrow$ remove ($B_{q,B-D,sam}(x,y)$)    ▶ Remove sample information
3:   $B_{q,edge}(x,y) \leftarrow$ edge ($B_{q,B-D}(x,y)$, canny)    ▶ Detect edge points with the canny operator
4:   $(x_e, y_e) \leftarrow$ arg ($B_{q,edge}(x,y) == 1$)    ▶ Record the coordinates of edge points
    **Step 2:** RANSAC algorithm
5:   for $k = 1$ to $K$ do
6:      $(x_{e1}, y_{e1}), (x_{e2}, y_{e2}), (x_{e3}, y_{e3}) \leftarrow (x_e, y_e)$    ▶ Randomly choose 3 edge points
7:      $(\hat{x}_{BD,q,k}, \hat{y}_{BD,q,k}), \hat{R}_{BD,q,k} \leftarrow (x_{e1}, y_{e1}), (x_{e2}, y_{e2}), (x_{e3}, y_{e3})$    ▶ Fit a circle
8:      $F_k = \sum_{x_e, y_e} p(x_e, y_e)$    ▶ Calculate the metric function
     end
9:   $(\hat{x}_{BD,q}, \hat{y}_{BD,q}), \hat{R}_{BD,q} \leftarrow k_{best} = \text{argmax}(F_k)$    ▶ Gain coarse circle center and radius
    **Step 3:** Denoise
10:   $(x_e, y_e)^u \leftarrow (x_e, y_e), (\hat{x}_{BD,q}, \hat{y}_{BD,q}), \hat{R}_{BD,q}$    ▶ Denoise with the coarse estimates
    **Step 4:** LS algorithm
11:   $(x_{BD,q}, y_{BD,q}), R_{BD,q} \leftarrow (x_e, y_e)^u$    ▶ Fit circle center and radius with LS algorithm
end

Next, estimate the full-pose parameters of LED array. In this step, we use the extracted circle centers and radii and the constructed physics model to estimate the six pose parameters of LED array. Mathematically, it can be described as

$$E(\Delta x, \Delta y, \theta_x, \theta_y, \theta_z, h) = \sum_{q=1}^{Q} \{[x_{BD,q} - x_{BD,q,c}]^2 + [y_{BD,q} - y_{BD,q,c}]^2\},$$
$$(\Delta x, \Delta y, \theta_x, \theta_y, \theta_z, h)^u = argmin[E(\Delta x, \Delta y, \theta_x, \theta_y, \theta_z, h)], \quad (14)$$

where $(x_{BD,q,c}, y_{BD,q,c})$ denotes the calculated coordinate using the six estimated parameters with our forward model. If the estimated parameters are close to the actual parameters, then the defined function $E(\Delta x, \Delta y, \theta_x, \theta_y, \theta_z, h)$ is close to zero. Hence, we can obtain the optimized pose parameters, $(\Delta x, \Delta y, \theta_x, \theta_y, \theta_z, h)^u$, by minimizing $E(\Delta x, \Delta y, \theta_x, \theta_y, \theta_z, h)$.

At last, calibrate the position of sub-spectra in the recovery process of FPM. With the estimated pose parameters, we can calculate the accurate position of each stitched spectrum in the Fourier domain. HR intensity and phase images can be obtained without degradation of imaging quality, regardless of how the LED array is placed.

### 3. Experiments

To validate our method experimentally, we compare the recovered results of one small segment ($128 \times 128$ pixels) in a USA-1951 resolution target with the conventional FPM, SC-FPM, and our method. One 21 × 21 LED array (2.5 mm spacing, 470 nm central wavelength with 20 nm bandwidth), an objective lens (NA = 0.1, Mag = 4), and a CMOS camera (FLIR, BFS-U3-200S6M-C, pixel size 2.4 μm) are used to build our FPM platform.

We first align the LED array with a 3D mechanical adjustment device that can achieve 2D lateral shifts and the in-plane rotation, and captured 441 LR intensity images to recover one HR intensity image as the reference, as shown in Fig. 5(a). To introduce pose misalignments, we shift, tilt and rotate the aligned LED array. Quantitatively, we treat the indications of the adjustment device as the nominal pose parameters of lateral shifts and in-plane rotation, and measure the height difference of the four corners of the LED array to calculate the nominal pose parameters of tilt. In the case of ($\Delta x = 2.000$ mm, $\Delta y = 2.000$ mm, $\theta_x = 3.224°$, $\theta_y = -2.149°$, $\theta_z = 0.000°$, $h = 92.00$ mm), the recovered intensity image with conventional FPM method is shown in Fig. 5(b1). The background is disturbed by wrinkle-shaped artifacts, and the features of all groups are also distorted and irresolvable. With the calibration of SC-FPM, an improved intensity image shown in Fig. 5(c1) is obtained, where the features of Group 8 can be resolved, while features of Group 9 are not clear due to the low contrast and the deteriorated background. What's more, the recovered pose parameters, ($\Delta x = 1.526$ mm, $\Delta y = 1.917$ mm, $\theta_z = -0.120°$, $h = 91.83$ mm), are not agree with the nominal parameter (the difference of the lateral shift along the $x$-axis is ~ 0.5 mm). After calibrating with our method, we get a high-quality HR intensity image, as shown in Fig. 5(d1), where all features can be resolved clearly and the contrast remains high. The estimated pose parameters, ($\Delta x = 1.998$ mm, $\Delta y = 1.988$ mm, $\theta_x = 2.540°$, $\theta_y = -1.671°$, $\theta_z = -0.218°$, $h = 91.42$ mm), are also consistent with the introduced parameters. Subsequently, we adjust the LED array to random position that contains an extreme in-plane rotation angle ($\theta_z$). The recovered intensity image with conventional FPM is shown in Fig. 5(b2). In addition to the decreased resolution, the stripes of the resolution target are rotated by an evident angle. Compared with Fig. 5(c1), the recovered image of SC-FPM shown in Fig. 5(c2) becomes worse. The features of Group 8, Element 6, have been irresolvable. Fortunately, with the calibration of the pose parameter of ($\Delta x = 2.168$ mm, $\Delta y = 0.348$ mm, $\theta_x = 3.421°$, $\theta_y = -2.258°$, $\theta_z = 5.562°$, $h = 91.44$ mm), the recovered image shown in Fig. 5(d2) is still comparable to the reference image.

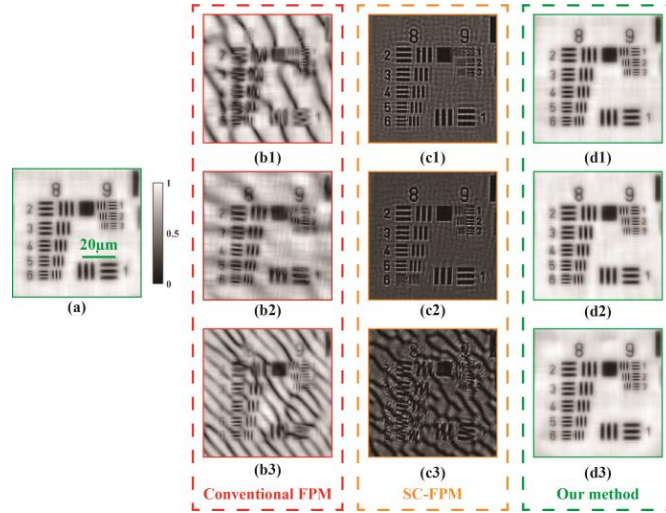

Fig. 5. Comparison of conventional FPM, SC-FPM, and our method. (b1)-(b3) are the recovered intensity images of conventional FPM with the nominal pose parameters of ($\Delta x = 2.000$ mm, $\Delta y = 2.000$ mm, $\theta_x = 3.224°$, $\theta_y = -2.149°$, $\theta_z = 0.000°$, $h = 92.00$ mm), a random position that contains an extreme misalignments of in-plane rotation ($\theta_z$), and ($\Delta x = 4.000$ mm, $\Delta y = 4.000$ mm, $\theta_x = 3.224°$, $\theta_y = -2.149°$, $\theta_z = 0.000°$, $h = 92.00$ mm), respectively. (c1)-(c3) are the corresponding recovered intensity images of SC-FPM. (d1)-(d3) are the corresponding recovered intensity images of our method.

Further, to validate the robustness and effectiveness of our method for calibrating remarkable misalignment, we deliberately adjust the LED array to an extremely misaligned pose to introduce the nominal parameters of ($\Delta x$ = 4.000 mm, $\Delta y$ = 4.000 mm, $\theta_x$ = 3.224°, $\theta_y$ = −2.149°, $\theta_z$ = 0.000°, $h$ = 92.00 mm). The recovered intensity image of conventional FPM is submerged by more wrinkle-shaped artifacts, as shown in Fig. 5(b3). With the calibration of SC-FPM, we get one intensity image shown in Fig.5 (c3), where the image quality seems lower than that of the uncalibrated image. The estimated pose parameters, ($\Delta x$ = −0.200 mm, $\Delta y$ = 1.913 mm, $\theta_z$ = 4.164°, $h$ = 93.10 mm), are also heavily skewed from the truth. The unstable performance of SC-FPM is caused by unsuitable initial step size and search direction, and it is difficult for unskilled users to optimize the initial parameters of SA algorithm. Thanks to the explicit physics model and effective algorithm, the recovered image with our method maintains high imaging quality, as shown in Fig. 5(d3). The recovered pose parameters, ($\Delta x$ = 4.075 mm, $\Delta y$ = 4.001 mm, $\theta_x$ = 3.325°, $\theta_y$ = -1.627°, $\theta_z$ = -0.242°, $h$ = 91.37 mm), are still close to the actual parameters, verifying that our method can achieve resolution- and contrast-invariant imaging for remarkable misalignment.

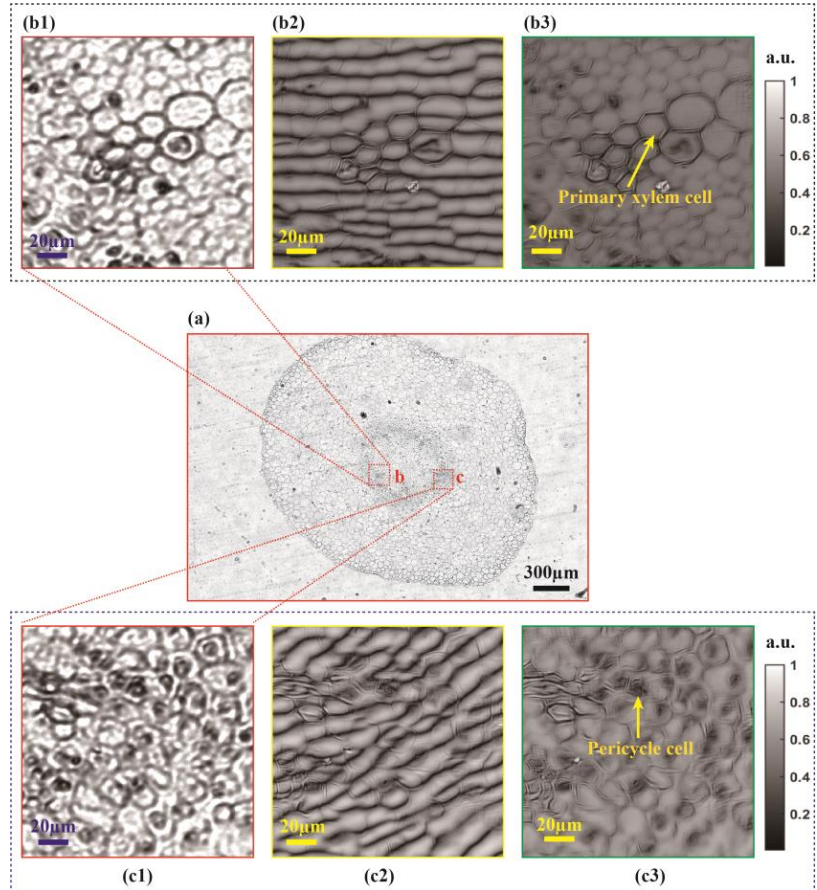

Fig. 6. Recovered results of two segments in a Vicia faba seedling root slide with conventional FPM and our method. (a) presents the captured full-FOV image of the sample under the illumination from the nominal central LED element. (b1)-(b3) show the enlargement of one small segment of (a), the recovered HR intensity images with conventional FPM and our method, respectively. (c1)-(c3) are the enlargement of another small segment of (a), the recovered HR intensity images with conventional FPM and our method, respectively.

In addition, we also demonstrate the effectiveness of our method by recovering a biological slide (Vicia faba seedling root). Similarly, we adjust the LED array to introduce unknown misalignments. Fig. 6(a) shows the captured full-FOV image of the sample under the illumination from the nominal central LED element. Fig. 6(b1) is a small segment of Fig. 6(a), with a resolution of ~ 4.7 μm, which is close to the size of the biological cell, resulting in the image being blurred. However, only by stitching captured LR images with conventional FPM method, we cannot get an ideal HR intensity image because of unmatched illumination angles. Instead, a distorted image coupled with noticeable artifacts is yielded, as shown in Fig. 6(b2). Using the B-D boundaries contained in captured images, we obtain the estimated pose parameters of ($\Delta x$ = -2.587 mm, $\Delta y$ = -0.271 mm, $\theta_x$ = 2.010°, $\theta_y$ = 1.731°, $\theta_z$ = -5.791°, $h$ = 91.97 mm), and an HR intensity image is recovered with the calibration of the parameters, as shown in Fig. (b3). Compared with Fig. 6(b2), the imaging quality is significantly improved and the evident artifacts are also removed. The image details, such as the cell mall of the primary xylem cells, can be clearly resolved as well. Fig. 6(c1) shows another small segment mainly consisting of pericycle cells. Although the recovered image without calibration is full of artifacts, as shown in Fig .6(c2), our method can still work well. The calibrated intensity image shown in Fig. 6(c3) has more distinct details, and the artifacts also have been eliminated.

## 4. Conclusion

We have demonstrated a physics-based and full-pose-parameter estimation method for correcting illumination angles in FPM systems. The effectiveness and robustness have been verified experimentally. Fundamentally, we analyze the structure of a typical FPM system to interpret the principle of forming arc-shaped B-D boundaries, and transform the problem of calibrating illumination angles from solving the position of LEDs to calculating the centers and radii of B-D boundaries. With the constructed model, we can solve for the six full-pose parameters to characterize a misplaced LED array accurately. Different from the pre-calibrated methods, our method is completely bind and can automatically remove random pose misalignments, unlocking the use of FPM by unskilled users. Further, compared with existing post-calibrated methods, our method is capable of modeling the LED array that is remarkably misplaced. Therefore, our method can facilitate the design, implementation, and application of concise and robust FPM platforms.

**Funding.** National Key Research and Development Program of China (2021YFC2202400); 111 Project (Medical Optics and Medical Imaging Overseas Expertise Introduction Center for Discipline Innovation) (B18005); Funding of Foundation Enhancement Program (2021-JCJQ-JJ-0823).

**Disclosures.** The authors declare no conflicts of interest.

**Data availability.** Data underlying the results presented in this paper are not publicly available at this time but may be obtained from the authors upon reasonable request.

## References

1. G. Zheng, R. Horstmeyer, and C. Yang, "Wide-field, high-resolution Fourier ptychographic microscopy," Nat. Photonics **7**(9), 739-745 (2013).
2. X. Ou, R. Horstmeyer, C. Yang, and G. Zheng, "Quantitative phase imaging via Fourier ptychographic microscopy," Opt. Lett. **38**(22), 4845-4848 (2013).
3. G. Zheng, C. Shen, S. Jiang, P. Song, and C. Yang, "Concept, implementations and applications of Fourier ptychography," Nat. Rev. Phy. **3**(3), 207-223 (2021).
4. S. A. Alexandrov, T. R. Hillman, T. Gutzler, and D. D. Sampson, "Synthetic Aperture Fourier Holographic Optical Microscopy," Phys. Rev. Lett. **97**(16), 168102 (2006).
5. J. R. Fienup, "Phase retrieval algorithms: a comparison," Appl. Opt. **21**(15), 2758-2769 (1982).
6. R. Horstmeyer, X. Ou, G. Zheng, P. Willems, and C. Yang, "Digital pathology with Fourier ptychography," Comput. Med. Imaging. Grap **42**, 38-43 (2015).
7. P. Song, S. Jiang, H. Zhang, X. Huang, Y. Zhang, and G. Zheng, "Full-field Fourier ptychography (FFP): Spatially varying pupil modeling and its application for rapid field-dependent aberration metrology," APL. Photonics **4**(5), 50802 (2019).
8. J. Chung, X. Ou, R. P. Kulkarni, and C. Yang, "Counting white blood cells from a blood smear using Fourier ptychographic microscopy," PLOS One **10**, e0133489 (2015).

9. R. Horstmeyer, J. Chung, X. Ou, G. Zheng, and C. Yang, "Diffraction tomography with Fourier ptychography," Optica **3**(8), 827-835 (2016).
10. L. Tian, and L. Waller, "3D intensity and phase imaging from light field measurements in an LED array microscope," Optica **2**(2), 104 (2015).
11. C. Zuo, J. Sun, J. Li, A. Asundi, Q. Chen, "Wide-field high-resolution 3D microscopy with Fourier ptychographic diffraction tomography," Opt. Laser. Eng. **128**, 106003(2020).
12. P. Thibault, M. Dierolf, O. Bunk, A. Menzel, and F. Pfeiffer, "Probe retrieval in ptychographic coherent diffractive imaging," Ultramicroscopy **109**, 338-343(2009).
13. H. M. L. Faulkner and J. M. Rodenburg, "Movable aperture lensless transmission microscopy: a novel phase retrieval algorithm," Phys. Rev. Lett. **93**, 023903 (2004).
14. S. Zhang, G. Zhou, Y. Wang, Y. Hu, and Q. Hao, "A Simply Equipped Fourier Ptychography Platform Based on an Industrial Camera and Telecentric Objective," Sensors **19**(22), 4913 (2019).
15. A. Zhou, W. Wang, N. Chen, E. Y. Lam, B. Lee, and G. Situ, "Fast and robust misalignment correction of Fourier ptychographic microscopy for full field of view reconstruction,", Biomed. Opt. Express **26**(18), 23661-23674 (2018).
16. C. Zheng, S Zhang, G, Zhou, Y. Hu, and Q. Hao, "Robust Fourier ptychographic microscopy via a physics-based defocusing strategy for calibrating angle-varied LED illumination," Biomed. Opt. Express **13**(3), 1581-1594 (2022).
17. D. Yang, S. Zhang, C. Zheng, G. Zhou, L. Cao, Y. Hu, and Q. Hao, "Fourier ptychography multi-parameunter neural network with composite physical priori optimization" Biomed. Opt. Express **13**(5), 2739-2753 (2022).
18. R. Eckert, Z. F. Phillips, and L. Waller, "Efficient illumination angle self-calibration in Fourier ptychography," Appl. Opt. **57**(19), 5434-5442 (2018).
19. J. Sun, Q. Chen, Y. Zhang, and C. Zuo, "Efficient positional misalignment correction method for Fourier ptychographic microscopy," Biomed. Opt. Express **7**(4), 1336 (2016).
20. A. Pan, Y. Zhang, T. Zhao, Z Wang, D. Dan, and B Yao, "System calibration method for Fourier ptychographic microscopy," J Biomed Opt. **22**(9), 1-11 (2017).
21. C. Zuo, J. Sun, and Q. Chen, "Adaptive step-size strategy for noise-robust Fourier ptychographic microscopy," Opt. Express **24**(18), 20724-20744 (2016)